%% file: manuscript.tex
\documentclass[mathpazo]{cicp}

\usepackage[utf8]{inputenc}
\usepackage{amsmath}
\usepackage{amsfonts}
\usepackage{bm}
\usepackage{fullpage}
\usepackage{subfig}
\usepackage[dvipsnames]{xcolor} 
\usepackage{tikz}
\usetikzlibrary{calc, positioning, shapes.arrows, arrows}
\DeclareMathOperator*{\argmin}{arg\,min}
\newcommand{\reynoldstress}{\boldsymbol{\tau}}
\newcommand{\reynoldstresscomp}{\tau}

\newcommand{\reynoldstressb}{\boldsymbol{b}}

\newcommand{\xstate}{\mathsf{x}}
\newcommand{\yobs}{\mathsf{y}}
\newcommand{\zout}{\mathsf{z}}
\newcommand{\Ccov}{\mathsf{C}}
\newcommand{\Hout}{\mathcal{H}}

\begin{document}
\title[Ensemble gradient for learning turbulence models]{
Ensemble gradient for learning turbulence models from indirect observations
}
\author[Michel\'en Str\"ofer, Zhang, and Xiao]{Carlos A. Michel\'en Str\"ofer\affil{**}\comma\affil{1}, 
Xin-Lei Zhang\affil{**}\comma\affil{1}\comma\affil{2}, and Heng Xiao\affil{1}\comma\corrauth}
\emails{{\tt hengxiao@vt.edu} (H.~Xiao)}
\address{\affilnum{1}\ Kevin T. Crofton Department of Aerospace and Ocean Engineering, Virginia Tech, Blacksburg, Virginia, USA\\
\affilnum{2}\ Arts et M\'etiers ParisTech, 8 Boulevard Louis XIV, 59046 Lille cedex, France\\
\affilnum{**}\ These authors contributed equally to this work.
}
\begin{abstract}
    Training data-driven turbulence models with high fidelity Reynolds stress can be impractical and recently such models have been trained with velocity and pressure measurements.  
    For gradient-based optimization, such as training deep learning models, this requires evaluating the sensitivities of the RANS equations. 
    This paper explores the use of an ensemble approximation of the sensitivities of the RANS equations in training data-driven turbulence models with indirect observations.
    A deep neural network representing the turbulence model is trained using the network's gradients obtained by backpropagation and the ensemble approximation of the RANS sensitivities. 
    Different ensemble approximations are explored and a method based on explicit projection onto the sample space is presented. 
    As validation, the gradient approximations from the different methods are compared to that from the continuous adjoint equations. 
    The ensemble approximation is then used to learn different turbulence models from velocity observations. 
    In all cases, the learned model predicts improved velocities. 
    However, it was observed that once the sensitivity of the velocity to the underlying model becomes small, the approximate nature of the ensemble gradient hinders further optimization of the underlying model. 
    The benefits and limitations of the ensemble gradient approximation are discussed, in particular as compared to the adjoint equations. 
\end{abstract}
\maketitle

\section{Introduction}
\label{sec:introduction}
    The Navier--Stokes equations fully describe the instantaneous velocity and pressure fields in fluid flows. 
    However the resolution required to capture the range of turbulence scales makes solving the Navier-Stokes equations computationally inaccessible for flows with high Reynolds number or complex geometries. 
    Instead, the Reynolds-averaged Navier--Stokes equations (RANS) are widely used in practice thanks to the relatively inexpensive computation required for their solution. 
    The RANS equations are a set of coupled partial differential equations (PDE) that describe the mean velocity ($\boldsymbol{u}$) and mean pressure ($p$) fields. 
    However, the RANS equations contain the unclosed Reynolds stress tensor $\reynoldstress$ which captures the effects of turbulence on the mean flow and requires modeling. 
    The incompressible steady RANS equations are
    \begin{equation}
        \begin{gathered}
            \boldsymbol{u\cdot\nabla u} - {\nu}\boldsymbol{\nabla^2u} +  \boldsymbol{\nabla\cdot}\reynoldstress + \nabla p - \boldsymbol{s} = \boldsymbol{0}, \\
            \boldsymbol{\nabla\cdot u} = 0,
        \end{gathered}
    \end{equation}
    where $\boldsymbol{s}$ are the external body forces. 
    Compactly, this can be written as $\mathcal{M}(\boldsymbol{u},p;\reynoldstress)=0$ where the Reynolds stress $\reynoldstress$ requires turbulence modeling. 
    
    Although widely used, RANS predictions are known to be inaccurate due to the lack of an accurate general turbulence model. 
    In particular, the widely used linear eddy viscosity models (LEVM) are known to be inaccurate even in simple flows of practical interest, including the inability to predict secondary flows in non-circular ducts~\cite{speziale1982turbulent}. 
    Eddy viscosity models are single-point closures that represent the Reynolds stress as a local function of the velocity gradient. 
    Non-linear eddy viscosity models (NLEVM) can capture more complex non-linear relations between the velocity gradient and the Reynolds stress, but existing NLEVM have not resulted in consistent improvement over LEVM. 
    This has led to an interest in developing data-driven turbulence models~\cite{duraisamy_turbulence_2018}. 
    Particularly, data-driven NLEVM~\cite{ling2016reynolds} have recently gained much attention. 
    
    Data-driven NLEVM have been typically trained with full field Reynolds stress data from high fidelity simulations. 
    It has been recently recognized, however, that the use of high fidelity Reynolds stress data for training can be impractical, which has led to the use of measurements derived from the velocity and pressure fields as training data~\cite{michelen2020Data,zhao_rans_2020}. 
    This allows for the use of more complex flows for which solutions of the Navier--Stokes equations are not feasible but for which experimental data is available. 
    Training the model using such data has the added complexity of requiring solving the RANS equations at each training step and, for gradient-based optimization, obtaining the gradient of the RANS equations. 
    In this work we explore the use of ensemble-based derivative approximations as an alternative to adjoint models for gradient-based training of data-driven turbulence models from indirect observations.

    \subsection{Data-driven eddy viscosity models trained with indirect observations}
    \label{sec:introduction:turb}
        The representation of the turbulence model and the training framework in this work are the same as in~\cite{michelen2020Data} except for the use of the ensemble gradient in place of the adjoint. 
        This framework is summarized here. 
        
        The Reynolds stress can be separated into isotropic and anisotropic components as
        \begin{equation}
            \reynoldstress = 2k\reynoldstressb - 2k\frac{1}{3}I, 
            \label{eq:reynoldsstress}
        \end{equation}
        where $\reynoldstressb$ is the normalized anisotropic (deviatoric) component of the Reynolds stress,  $k$ is the turbulent kinetic energy, and $I$ is the second rank identity tensor. 
        An eddy viscosity turbulence model is an invariant mapping from the mean flow velocity gradient to the Reynolds stress tensor, $\nabla \boldsymbol{u}\mapsto\reynoldstress$. 
        Any such mapping can be expressed using the integrity basis representation~\cite{pope1975more} as 
        \begin{equation}
        \begin{gathered}
            \reynoldstressb = \sum_{i=1}^{10}g^{(i)}T^{(i)}, \\
            g^{(i)} = g^{(i)}(\theta_1,\dots,\theta_{5}), 
        \end{gathered} \label{eq:reynoldsstressb}
        \end{equation}
        $T^{(i)}$ are the tensor functions in the integrity basis, $g^{(i)}$ are the scalar coefficient functions, and $\theta_j$ are the scalar invariants of the input velocity gradient tensor. 
        The full list of basis tensors and input invariants are presented in~\cite{pope1975more} and the linear and quadratic terms used here are 
        \begin{equation}
            \begin{array}{ll}
            \displaystyle T^{(1)}=S, \quad & T^{(2)}=SR-RS,\\[3pt] 
            \displaystyle T^{(3)}=S^2-\frac{1}{3}\left\{S^2\right\}I, \quad &
            T^{(4)}=R^2-\frac{1}{3}\left\{R^2\right\}I, 
            \end{array}
            \label{eq:tensorbasis}
        \end{equation}
        and 
        \begin{equation}
          \theta_1=\left\{S^2\right\},\quad\theta_2=\left\{R^2\right\}, 
          \label{eq:scalarbasis}
        \end{equation}
        where $S$ and $R$ are the non-dimensionalized symmetric and antisymmetric components of the velocity gradient tensor. 
        Equations~\eqref{eq:reynoldsstress}--\eqref{eq:scalarbasis} together with transport equations for the turbulent kinetic energy $k$ and at least one more turbulent scale---to obtain the turbulence time scale for non-dimensionalizing the velocity gradient---constitute a complete eddy visocity model. 
        Data-driven eddy viscosity models retain the turbulent scales transport equations from a traditional model but learn the closure form, i.e. the functions $g^{(i)}(\theta_1,\dots,\theta_{5})$, from data. 
        Here, a deep neural network is used to represent this mapping, $\boldsymbol{\theta}\mapsto\boldsymbol{g}$. 
        
        \begin{figure}
            \centering
            \input{figure_1.tikz}
            \caption{Schematic of the training framework adapted from~\cite{michelen2020Data}. 
            For any value of the network's parameters, $\boldsymbol{w}$, the gradient of the objective function, $J$, can be obtained by combining the gradients from backpropagation of the neural network and the gradient of the PDEs (e.g. RANS).
            The PDE gradients (highlighted in red/dark grey) are obtained here by using ensemble approximations, which is in contrast to that in~\cite{michelen2020Data} where it is obtained by solving the RANS adjoint equations.
            }
            \label{fig:overview}
        \end{figure}
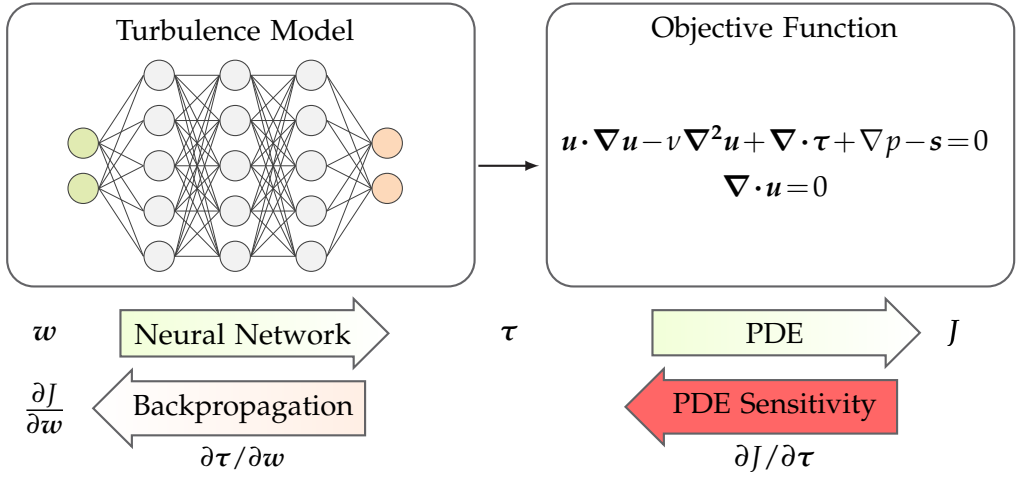
        
        Training the neural network with indirect observations, i.e. with velocity and pressure quantities rather than with Reynolds stress, requires derivative information for both the neural network and the RANS equations. 
        The overall training framework is summarized in Figure~\ref{fig:overview}. 
        The gradient of the neural network outputs with respect to its trainable parameters are obtained using backpropagation, an efficient reverse mode automatic differentiation algorithm for deep neural networks. 
        The cost function $J$ compares the predicted and measured quantities and requires solving the RANS equations. 
        The gradient of the cost function with respect to the trainable parameters $\boldsymbol{w}$ is given by the chain rule as 
        \begin{equation}
            \frac{\partial J}{\partial \boldsymbol{w}} = \frac{\partial J}{\partial \reynoldstress} \frac{\partial \reynoldstress
            }{\partial \boldsymbol{w}}. 
        \end{equation}
        Here the sensitivity of cost function to the Reynolds stress $\partial J/\partial\reynoldstress$ is obtained via an ensemble approximation rather than by solving the adjoint equations as in~\cite{michelen2020Data}.

    \subsection{Differentiation of physical models}
    \label{sec:introduction:gradient}
        Training data-driven models with indirect observations poses a challenge to computing the gradient of the cost function as this now requires obtaining the sensitivities of the RANS equations. 
        Three options are (i) using gradient-free optimization to avoid the gradient calculation, (ii) solving the adjoint equations to obtain the exact gradient, or (iii) approximating the gradient from multiple evaluations of the RANS model. 
        
        \paragraph{Gradient-free optimization} 
        Gradient-free optimization updates an ensemble of states based on a heuristic other than gradient-descent, such as natural selection in genetic algorithms. 
        Zhao et al.~\cite{zhao_rans_2020} used gene expression programming to train a model in a gradient-free manner. 
        Although this can be an effective approach, using the gradient information is more efficient and should be used if available~\cite{audet2016blackbox}. 
        In particular, the recent success of deep learning methods~\cite{lecun2015deep} is in large part do to the use of novel gradient-based optimization methods.
        
        \paragraph{Adjoint equations} 
        Solving the adjoint equations is an efficient method for obtaining the exact derivative information~\cite{othmer2008continuous,he2018data, sabater2021gradient}. 
        Both the continuous adjoint~\cite{michelen2020Data} and the discrete adjoint~\cite{holland2019towards} have been used to obtain the sensitivities of the RANS equations and train a deep neural networks using gradient-based optimization.
        The adjoint equations, however, can suffer from known instabilities making it difficult to converge~\cite{oriani2016alternative}. 
        More importantly, the adjoint method is intrusive requiring significant effort, e.g. to derive continuous adjoint equations for a new problem or to implement the discrete adjoint on existing large codes via automatic differentiation tools.
        
        \paragraph{Numerical approximations} In this work we explore the use of approximate derivatives using ensemble methods in place of the exact derivatives from the continuous or discrete adjoint. 
        Like the gradient-free approach this requires multiple evaluations of the RANS equations and treats the model as a black box but like the adjoint methods it provides a gradient that can be used in gradient-based optimization. 
        Different types of numerical approximations of model derivatives have been devised, including the finite difference~\cite{conn2009introduction}, simplex gradient~\cite{nocedal2006numerical}, simultaneous perturbation stochastic approximation (SPSA)~\cite{spall1992multivariate}, and ensemble-based optimization (EnOpt)~\cite{chen2009efficient}.
        These methods differ on the number and selection of samples and how these are used to estimate the gradient.
        Specifically, finite difference uses the same number of samples, $N$, as dimensions, $D$, and perturbs one orthogonal direction at a time with a fixed perturbation.
        The simplex gradient method uses a set of $N=D+1$ samples that are affinely independent and estimates the gradient information based on the relationship between the samples and the centroid.
        Simultaneous perturbation stochastic approximation uses $N=2$ samples by simultaneously perturbing all model parameters to estimate the gradient.
        Further, a modified SPSA~\cite{li2011uncertainty} or stochastic Gaussian search direction (SGSD) method draws multiple ($N<D$) samples from a Gaussian distribution and uses the expectation of estimated gradient as the downhill direction.
        The ensemble gradient method used here also uses $N<D$ random samples from a Gaussian process and has the benefit of significantly reducing the number of model evaluations as compared to finite difference or simplex gradient. 
        The ensemble gradient is more robust and efficient for finding the steepest direction than other approximate gradient methods and can provide conditioned realizations based on random maximum likelihood~\cite{li2011uncertainty}.
        The connections between the ensemble-based optimization and other methods is discussed in~\cite{do2013theoretical}, where it is shown that the ensemble-based optimization is equivalent to a preconditioned simplex gradient and to the second-order SPSA with Gaussian sampling. 
        The ensemble-based gradient is also employed implicitly in ensemble-based data assimilation techniques such as ensemble Kalman filtering~\cite{evensen2003ensemble} and ensemble maximum randomized likelihood method~\cite{gu2007iterative}.
        Ensemble-based data assimilation performs an implicit optimization, but it essentially uses an explicit approximate stochastic gradient based on the ensemble realizations~\cite{kovachki2019ensemble, do2013theoretical}.

    \subsection{Contribution of present work}
    \label{sec:introduction:contrib}
        In this work we demonstrate the use of the ensemble gradient approximation to train a data-driven turbulence model represented by a deep neural network. 
        Specifically, we combine the ensemble approximations of the RANS sensibilities with the gradients of the neural network to obtain a fully differentiable training framework. 
        While we use it for turbulence modeling we point out that this differentiable \emph{deep neural netowrk plus PDEs} framework could be used for differentiable learning of any physical model (represented by the neural network) from indirect observations, i.e. observations requiring the solutions of additional models (PDEs). 
        The ensemble gradient represents a non-intrusive method for obtaining gradients that can treat the PDEs as a black box model. 
        Additionally the ensemble gradient can be robust to non-differentiable or noisy problems as long as there is a well behaved larger trend. 
        This can be important, for example, for the sensitivity of high fidelity turbulent flows whose chaotic nature makes the adjoint method ineffective~\cite{blonigan2018multiple}.
        Here, we use both the EnOpt method~\cite{chen2009efficient} and a method derived here which is based on explicit projection onto the samples. 
        The new method performs just as well but is based on a different heuristic. 
        
        The rest of the paper is organized as follows.
        Section~\ref{sec:method} presents the different ensemble-based methods, their interpretations, and specialization to the turbulence modeling problem. 
        Section~\ref{sec:results} presents the results, including a comparison of the computed gradients between different methods and the adjoint as well as two test cases of learning turbulence models from velocity observations. 
        Finally, Section~\ref{sec:conclusion} concludes the paper.

\section{Ensemble Gradient}
\label{sec:method}
    The ensemble gradient approximation is characterized by the use of $N$ random samples, with $N$ less than the dimensions of the problem. 
    The samples are chosen from a Gaussian process with assumed covariance kernel. 
    This section present different approaches for using ensemble approximation for the gradient in the context of optimization. 
    The direct implementation of gradient descent with ensemble gradient is presented first and the reason why it fails is discussed. 
    Next, the common approach of preconditioning the gradient descent with the state covariance is presented. 
    Lastly, the method used in this work, which is based on direct projection onto the subspace spanned by the samples, is presented.
    This section concludes with a summary of the application of these different methods to the problem of training data-driven NLEVM with indirect observations. 
    Numerical comparison of these methods will be shown in Section~\ref{sec:results:validation}. 
    
    \subsection{Direct ensemble gradient}
    \label{sec:method:naive}
        The cost function to be minimized is the least squares discrepancy between the observations $\yobs$ and predictions $\zout$, given as 
        \begin{gather}
            J(\xstate) = \frac{1}{2}\lVert \zout - \yobs \rVert^2_{\Ccov_\yobs^{-1}}, \label{eq:cost}\\
            \zout = \Hout(\xstate), 
        \end{gather}
        where $\xstate$ is the state vector to be optimized, $\lVert\cdot\rVert^2_W$ indicates the L-2 norm weighted by the matrix $W$, $\yobs$ is the vector of measurement data, and $\Ccov_\yobs$ is the measurement covariance matrix representing measurement uncertainties. 
        For a given state $\xstate$ the model predictions $\zout$ are the state mapped to observation space by means of the non-linear observation operator $\Hout$ that maps from state space to observation space. 
        For gradient descent optimization the state is updated as
        \begin{equation}
            \xstate_{n+1} = \xstate_n - \alpha_n\nabla J(\xstate_n), \label{eq:gradDescent}
        \end{equation}
        where $\alpha$ controls the step size and can be a fixed scalar or determined by line search at each step $n$. 
        The gradient of the cost function in Equation~\ref{eq:cost} is 
        \begin{equation}
            \nabla J(\xstate) = (\Hout^\prime(\xstate))^\top \Ccov_\yobs^{-1}\left(\Hout(\xstate) - \yobs\right), \label{eq:gradorg}
        \end{equation}
        where $\Hout^\prime(\xstate)$ is the sensitivity matrix of the vector $\Hout(\xstate)$ with respect to the state $\xstate$. 
        The matrix $\Ccov_\yobs$ and vector $\yobs$ are fixed while the vector $\Hout(\xstate)$ is obtained by solving the forward (e.g. RANS) model. 
        The sensitivity matrix $\Hout^\prime(\xstate)$ is estimated using an ensemble gradient approximation as
        \begin{equation}
            \Hout^\prime(\xstate) \approx [\Hout(\xstate^{(i)}) - \overline{\Hout(\xstate)}][\xstate^{(i)} - \bar{\xstate}]^{+} = \Delta_{\zout}\Delta_\xstate^{+}, \label{eq:gradDirect}
        \end{equation}
        where for a quantity $\phi$ the matrix $\Delta_\phi=[\phi^{(1)} - \bar{\phi},\dots,\phi^{(N)} - \bar{\phi}]$ consists of the mean-subtracted samples, $\phi^{(i)}$ indicates the $i$\textsuperscript{th} sample, $\bar{\phi}$ is the ensemble mean, and the superscript $+$ indicates the pseudoinverse. 
        However, the inverse problem described by Equation~\eqref{eq:cost} is ill-posed in that multiple states $\xstate$ can result in $\Hout(\xstate)=\yobs$ or at least $\Hout(\xstate)\approx\yobs$ up to high accuracy. 
        This inherent ill-posedness can result in a cost function and sensitivity matrix with frequent, abrupt changes for small changes in the state $\xstate$. 
        Even for well-posed problems the use of the pseudoinverse of the ensemble matrix can lead to noisy approximate gradients, as will be demonstrated later. 
        For this reason the direct ensemble gradient is not a suitable option for gradient descent optimization or model learning as pursued in this work.

    \subsection{Preconditioning with state covariance}
    \label{sec:method:covariance}
        Pre-multiplying the gradient $\nabla J(\xstate)$ by the state covariance $\Ccov_{\xstate}$ is a common method for preconditioning of the gradient descent optimization, as 
        \begin{equation}
            \xstate_{n+1} = \xstate_n - \alpha_n\Ccov_{\xstate} \nabla J(\xstate_n), \label{eq:gradDescentP}
        \end{equation}
        referred to as EnOpt~\cite{chen2009efficient}.
        This correspond to steepest descent rather than gradient descent on a discrete vector space with inner-product defined by the covariance $\Ccov_{\xstate}$~\cite{tarantola2005inverse_ch6}. 
        The use of the state covariance as a preconditioner for the gradient descent is discussed in more detail in Appendix~\ref{app:premultiply}.
        The ensemble approximation of the state covariance is given as  
        \begin{equation}
            \Ccov_{\xstate} \approx \frac{1}{N}\Delta_{\xstate}\Delta_\xstate^\top. 
        \end{equation}
        The product of the state covariance and the forward model sensitivity, which arises when substituting Equation~\eqref{eq:gradorg} into Equation~\eqref{eq:gradDescentP}, can be approximated using the ensemble as 
        \begin{align}
        \begin{split}
            \Ccov_{\xstate}\Hout^\prime(\xstate) &\approx \frac{1}{N}\Delta_{\xstate}\Delta_\xstate^\top (\Delta_{\yobs}\Delta_\xstate^{+})^\top\\
            &= \frac{1}{N}\Delta_{\xstate}\Delta_\xstate^\top(\Delta_\xstate^\top)^{+}\Delta_{\yobs}^\top\\
            &= \frac{1}{N}\Delta_{\xstate}\Delta_{\yobs}^\top\\
            &\approx \Ccov_{\xstate \zout}, 
        \end{split}
        \label{eq:Cxh}
        \end{align}
        where $\Ccov_{\xstate \zout}$ is the cross-covariance between the state and model outputs. 
        This formulation avoids taking the pseudoinverse of an ensemble matrix, which in general can be ill-conditioned. 
        Empirical results show that while the ensemble gradient in Equation~\eqref{eq:gradDirect} is noisy, the ensemble cross-covariance $\Ccov_{\xstate \yobs}$ captures the correct qualitative correlations~\cite{chen2012ensemble}. 
        Finally, the update scheme in Equation~\eqref{eq:gradDescentP} becomes
        \begin{equation}
            \xstate_{n+1} = \xstate_n - \alpha_n \Ccov_{\xstate \zout} \Ccov_\yobs^{-1}\left(\Hout(\xstate_n) - \yobs\right),
        \end{equation}
        where $\Ccov_{\xstate \zout}$ is approximated with the ensemble, based on Equation~\eqref{eq:Cxh} as 
        \begin{equation}
            \Ccov_{\xstate \zout}\approx\frac{1}{N}\Delta_{\xstate}\Delta_{\zout}^\top.
        \end{equation}

    \subsection{Projection to subspace}
    \label{sec:method:projection}
        The method used here is based on explicit projection of the state onto the subspace spanned by the samples in the ensemble, similar to the formulation in the ensemble-variational (EnVar) data assimilation method~\cite{liu2008ensemble}. 
        The ill-posedness of the inverse problem in Equation~\eqref{eq:cost} is alleviated by means of dimensionality reduction and regularized projection. 
        The state is expressed as $\xstate=\bar{\xstate}+\Delta_{\xstate}\beta$ where $\Delta_{\xstate}$ is the matrix of mean subtracted samples and the vector $\beta\in\mathbb{R}^N$ is the new state with reduced dimensions. 
        That is, a state $\xstate$ is expressed as the ensemble mean plus a linear combination of mean-subtracted samples. 
        The cost function in Equation~\eqref{eq:cost} now has gradient, with respect to the new state $\beta$, given as
        \begin{align}
        \begin{split}
            \nabla_{\beta} J(\xstate(\beta)) &= \Delta_\xstate^\top(\Hout^\prime(\xstate))^\top \Ccov_\yobs^{-1}\left(\Hout(\xstate(\beta))-\yobs\right) \\
            &\approx \Delta_\xstate^\top (\Delta_{\zout}\Delta_\xstate^{+})^\top \Ccov_\yobs^{-1}\left(\Hout(\xstate(\beta))-\yobs\right) \\
            &= \Delta_{\zout}^\top \Ccov_\yobs^{-1}\left(\Hout(\xstate(\beta))-\yobs\right),
        \end{split} \label{eq:dJdbeta}
        \end{align}
        where the model discrepancies $\Delta_{\zout}$ are obtained from the ensemble. 
        While this gradient could be used to optimize the new state $\beta$, for training the turbulence model the gradient of the cost function with respect to the original state $\xstate$ is required. 
        For this, we seek the gradient of the new state $\beta$ with respect to the original state $\xstate$. 
        For a state $\xstate$, the new state $\beta$ can be obtained by projecting $\xstate-\bar{\xstate}$ onto the mean-subtracted samples as
        \begin{equation}
            \beta_i = \frac{\left<\xstate-\bar{\xstate}, \Delta_\xstate^{(i)}\right>_W}{\left<\Delta_\xstate^{(i)},\Delta_\xstate^{(i)}\right>_W},
        \end{equation}
        or
        \begin{equation}
            \beta = (\Delta_\xstate^\top W \Delta_{\xstate})^{-1}\Delta_\xstate^\top W (\xstate-\bar{\xstate}),
        \end{equation}
        where $W$ is the weight matrix in the definition of the vector inner product $\left< \cdot \right>$. 
        E.g. if $\xstate$ is a discretized field, $W$ is the diagonal matrix containing cell volumes. 
        The gradient is then
        \begin{equation}
            \nabla\beta(\xstate) = (\Delta_\xstate^\top W \Delta_{\xstate})^{-1}\Delta_\xstate^\top W. 
        \end{equation}
        It is noted that the matrix $\Delta_\xstate^\top W \Delta_{\xstate}$ is not related to the state covariance $\Ccov_{\xstate}\approx\frac{1}{N}\Delta_{\xstate}\Delta_\xstate^\top$ and is often ill-conditioned as the samples are randomly drawn and not necessarily orthogonal. 
        A convenient approach to conditioning the problem is Tikhonov regularization which results in 
        \begin{equation}
            \beta = (\Delta_\xstate^\top W \Delta_{\xstate} + \lambda \mathsf{I})^{-1}\Delta_\xstate^\top W (\xstate-\bar{\xstate})
        \end{equation}
        and consequently
        \begin{equation}
            \nabla\beta(\xstate) = (\Delta_\xstate^\top W \Delta_{\xstate} + \lambda \mathsf{I})^{-1}\Delta_\xstate^\top W, \label{eq:dbetadx}
        \end{equation}
        where $\mathsf{I}$ is the identity matrix and $\lambda$ is the Lagrange multiplier of the constraint which can typically be very small. 
        The details are presented in Appendix~\ref{app:reg}. 
        
        The desired gradient can be obtained by using the chain rule to combine Equation~\eqref{eq:dJdbeta} and Equation~\eqref{eq:dbetadx} as
        \begin{equation}
            \nabla J(\xstate) = \left( (\Delta_\xstate^\top W \Delta_{\xstate} + \lambda \mathsf{I})^{-1}\Delta_\xstate^\top W \right)^\top \left( \Delta_{\zout}^\top \Ccov_\yobs^{-1}\left(\Hout(\xstate)-\yobs\right) \right),
        \end{equation}
        where $\Delta_{\zout}$ and $\Delta_{\xstate}$ are both obtained from the ensemble. 
        The ensemble of mean-subtracted states $\Delta_{\xstate}$ could be fixed throughout the optimization, and only the ensemble of model outputs $\Delta_\zout$ be updated using the fixed $\Delta_{\xstate}$ but updated mean $\bar{\xstate}=\xstate_n$.  
        Here, however, we follow the common practice in EnVar~\cite{mons2016reconstruction} and resample the ensemble state at each iteration. 
        Similarly, we use the true mean $\bar{\xstate}=\xstate_n$ and $\overline{\Hout(\xstate)}=\Hout(\bar{\xstate})$ rather than the ensemble mean for the ensemble discrepancies and denote these true means by $\widetilde{\Delta}_{\xstate}$ and $\widetilde{\Delta}_{\Hout}$, respectively. 
        Finally the gradient is given by 
        \begin{equation}
            \nabla J(\xstate) = \left( (\widetilde{\Delta}_\xstate^\top W \widetilde{\Delta}_{\xstate} + \lambda \mathsf{I})^{-1}\widetilde{\Delta}_\xstate^\top W \right)^\top \left(\widetilde{\Delta}_{\zout}^\top \Ccov_\yobs^{-1}\left(\Hout(\xstate)-\yobs\right)\right). 
        \end{equation}

    \subsection{Ensemble gradient for the RANS equations}
    For the present case of training a turbulence model from indirect observations, the cost function is given by Equation~\eqref{eq:cost} with the state consisting of the discretized values of the Reynolds stress $x=\reynoldstress$ at each cell, and the forward model $\Hout$ consisting of the composition of the RANS equations, which map Reynolds stress to velocity and pressure $\reynoldstress\mapsto (\boldsymbol{u},p)$, and an observation operator that maps velocity and pressure fields to observations (e.g. sparse sampling, surface integration). 
    Regardless of which method is used, the ensemble gradient requires solving the RANS equations for the neural network predicted Reynolds stress, i.e. obtaining $\Hout(\reynoldstress)$, as well as creating an ensemble of Reynolds stress fields $\Delta_{\reynoldstress}$ and solving the RANS equations for each sample in the ensemble to obtain the ensemble of model predictions $\Delta_{\zout}$. 
    The gradient used for each of the three methods are summarized in Table~\ref{tab:gradients}. 
    For the projection method, since the state is a discretization of a field, the matrix $W$ defining the inner product is the diagonal matrix $D_V$ with cell volumes along the diagonal. 
    
    \begin{table}[!htb]
        \centering
        \renewcommand{\arraystretch}{1.5}
        \begin{tabular}{c|c}
          \hline
            method & $\partial J/\partial \reynoldstress$ \\
            \hline
            direct & $(\Delta_{\zout}\Delta_{\reynoldstress}^{+})^\top \Ccov_\yobs^{-1}\left(\Hout(\reynoldstress) - \yobs\right)$ \\
            preconditioned & $\frac{1}{N}\Delta_{\reynoldstress}\Delta_{\zout}^\top \Ccov_\yobs^{-1}\left(\Hout(\reynoldstress) - \yobs\right)$ \\
            projection & $\left( (\widetilde{\Delta}_{\reynoldstress}^\top D_V \widetilde{\Delta}_{\reynoldstress} + \lambda \mathsf{I})^{-1}\widetilde{\Delta}_{\reynoldstress}^\top D_V \right)^\top \left(\widetilde{\Delta}_{\zout}^\top \Ccov_\yobs^{-1}\left(\Hout(\reynoldstress)-\yobs\right)\right)$ \\
              \hline
        \end{tabular}
        \renewcommand{\arraystretch}{1.0}
        \caption{Ensemble gradient approximations of the sensitivites of the cost function to the Reynolds stress using the different methods. All quantities of the form $\Delta_\phi$ or $\widetilde{\Delta}_\phi$ are ensemble matrices.}
        \label{tab:gradients}
    \end{table}
    
    Creating the ensemble of states requires perturbing the Reynolds stress tensor, which poses the question of how to best perturb a tensor field. 
    This has been done in previous works by perturbing the turbulence kinetic energy (magnitude), eigenvalues (shape), and eigenvectors (direction) of the Reynolds stress tensor~\cite{xiao2016quantifying}. 
    The formulation of the Reynolds stress in Equation~\eqref{eq:reynoldsstress}, however, provides a convenient way of creating the ensemble by perturbing the predicted coefficients $g^{(i)}$. 
    This is done by sampling the coefficients $g^{(i)}$ from a Gaussian process and using Equations~\eqref{eq:reynoldsstress} and~\eqref{eq:reynoldsstressb} to reconstruct the Reynolds stress.  
    One challenge, however, is that a covariance matrix must be specified for each coefficient function. 
    Simply using the same covariance for all functions is not adequate because of the large difference in magnitude between them. 
    The order of magnitude of the coefficients are not known a priori and can change during the optimization process. 
    To address this challenge, the samples are created from Gaussian processes $g^{(i),(j)}\sim GP(g^{(i)},\Ccov_g^{(i)})$, where each covariance kernel $\Ccov_g^{(i)}$ shares the same correlation kernel but has a standard deviation field proportional to the norm of $g^{(i)}$. 
    The correlation kernel $K$ used is a square exponential, and the covariance kernel is given by 
    \begin{align}
        \Ccov_g^{(i)} &= \alpha^2 \lVert g^{(i)}\rVert^2_{D_V} K, \\
        K(x_i, x_j) &= \exp\left( -\frac{1}{2}\frac{\lVert x_i-x_j\rVert^2}{l^2} \right),
    \end{align}
    where $x_i$ and $x_j$ are the spatial coordinates of cell $i$ and $j$, $l$ is the correlation length, and $\alpha$ is the ratio of the standard deviation to the magnitude of $g^{(i)}$.

\section{Results}
\label{sec:results}
    The ensemble gradient is first validated by comparing the ensemble gradient, using different methods, to the gradient obtained from the adjoint equations for a turbulent channel flow.  
    The ensemble gradient is then used in place of the adjoint for learning 
    turbulence models from synthetic full field velocities in two cases, similar to those in~\cite{michelen2020Data}. 
    The use of synthetic data provides a ground truth to which to compare the learned models. 
    The first case is a turbulent channel flow with a linear closure model. 
    The second case is flow through a square duct with a quadratic model. 
    For both cases, a neural network with $10$ hidden layers and $10$ neurons per hidden layer is used. 
    A ReLU activation is used for the hidden layers and a linear activation for the output layer.
    The training is done with the ADAM algorithm, a common gradient-based training algorithm for deep neural networks. 

    \subsection{Validation}
    \label{sec:results:validation}
        For validation, the fully developed turbulent channel flow is used. 
        The RANS simulation domain includes the bottom half of the channel, with a symmetry boundary condition at the mid-channel, and a discretization of $50$ cells of equal sizes. 
        The Reynolds number, based on bulk velocity $u_b$ and channel half height $h$, is $10,000$, and the linear $k$--$\omega$ model is used for the truth.  
        The cost function is the full field discrepancy of the velocity $u_{x}$, i.e. 
        \begin{equation}
            J(\reynoldstress) = \lVert u_{x}(\reynoldstress) - u_x^* \rVert^2_{D_V},
        \end{equation}
        where $u_x^*$ is the true velocity field and $D_V$ contains the cell volumes in the diagonal. 
        Two validation studies are presented where the different ensemble gradients are compared to the gradient from solving the adjoint equations. 
        
\begin{figure}[!htb]
            \centering \includegraphics[width=0.8\textwidth]{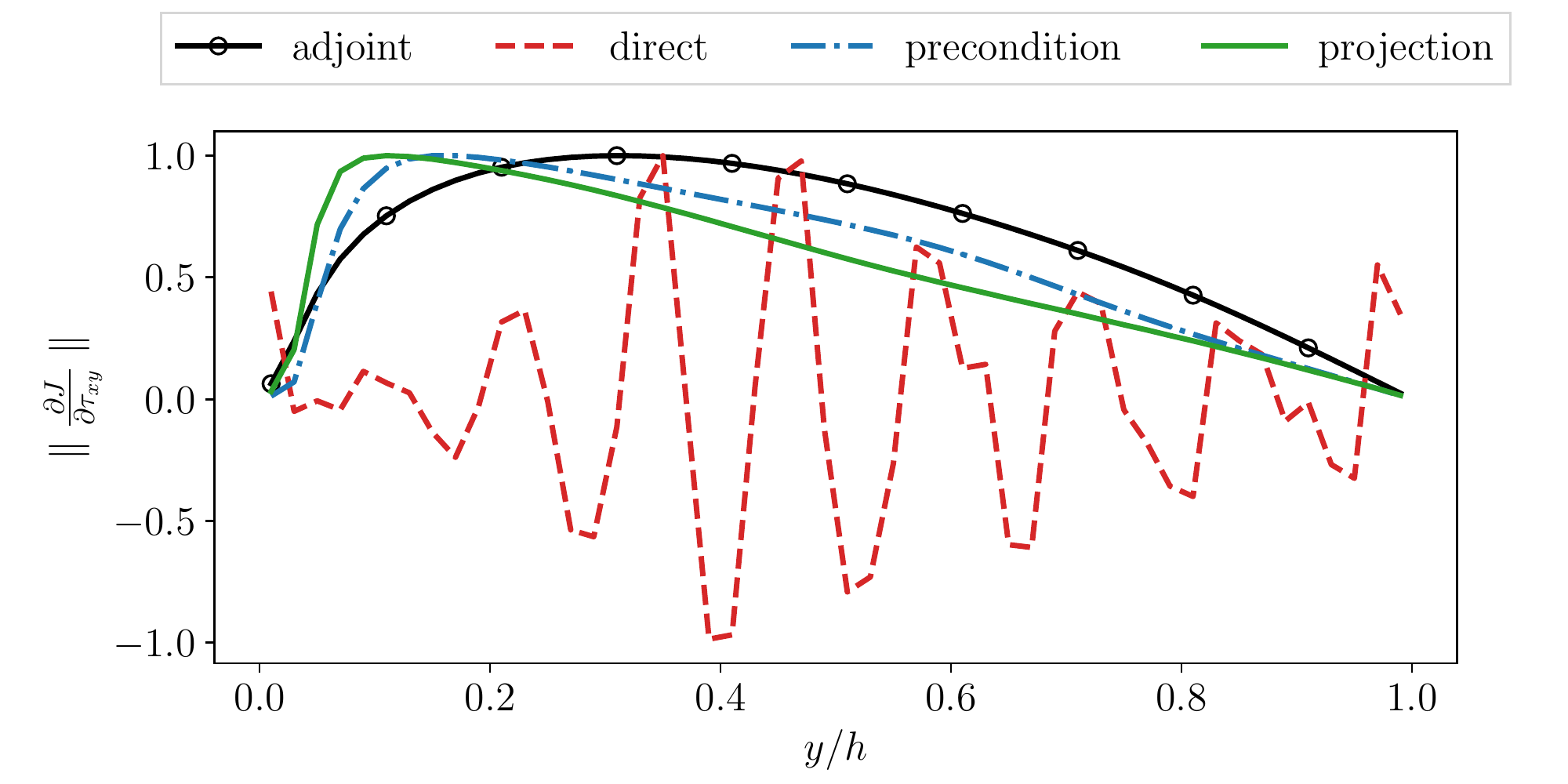}
            \caption{Comparison of sensitivity $\frac{\partial J}{\partial \reynoldstresscomp_{xy}}$ among adjoint method, direct ensemble method, precondition method and projection method. The sensitivity is normalized by $\text{max}\left(\frac{\partial J}{\partial \reynoldstresscomp_{xy}}\right)$ to keep the maximum value as $1$}
            \label{fig:comp_sensivity}
        \end{figure}

        First, the sensitivity of the cost function to the Reynolds stress $\partial J/\partial\reynoldstress$ is compared between the different methods at $\tau=0$, corresponding to laminar velocity. 
        The ensemble sensitivities are approximated with $20$ samples. 
        The results are shown in Figure~\ref{fig:comp_sensivity}, where it can be seen that the direct ensemble method gives a noisy estimate on the sensitivity. 
        In contrast, the ensemble method with either the projection method or the precondition method can provide smooth estimation of the sensitivity and give the same gradient direction as the adjoint method. 
        This suggests that either the projection or preconditioned ensemble methods can be used for gradient-based training. 
        It is noted that the exact values of the gradient are not expected to be the same, but as long as it has the right sign and has the correct zero any estimate of the gradient can be used for training. 
        One reason for the discrepancy is that the adjoint equations give the gradient of the Laplacian function that includes the cost function and the RANS constraints~\cite{michelen2021machine}. 
        This is in addition to the other approximations in both the adjoint and ensemble methods. 
        
\begin{figure}[!htb]
            \centering            
            \includegraphics[width=0.5\textwidth]{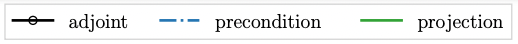}\\
            \includegraphics[width=0.6\textwidth]{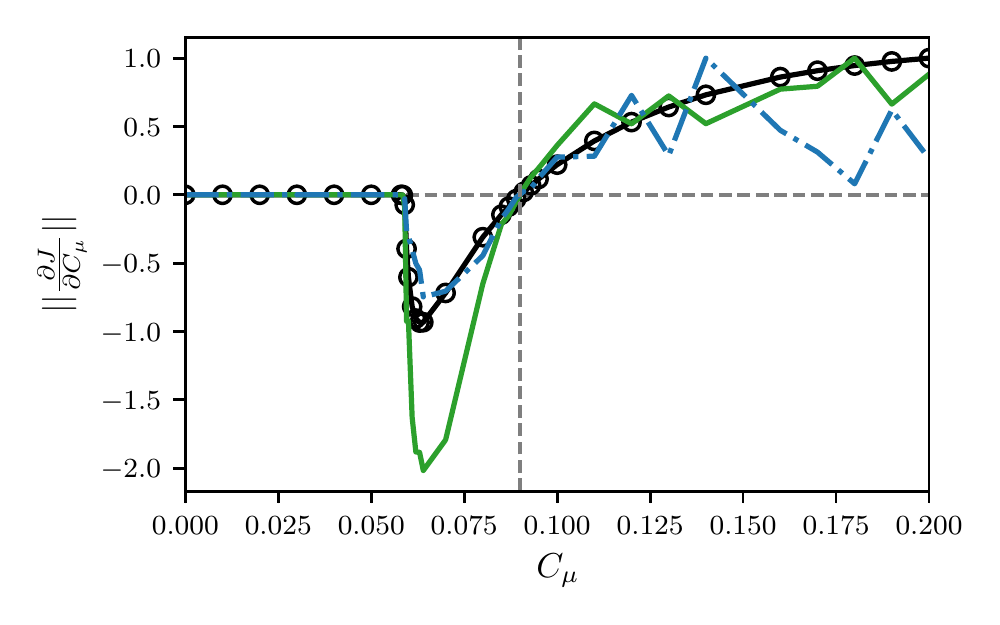}
            \caption{Comparison of the gradient of the cost function for the simplified one-parameter turbulence model using different methods. The gradients are normalized based on their maximum value. The vertical dashed line indicate the synthetic truth $C_\mu = 0.09$.}
            \label{fig:comp_gradient}
           
        \end{figure}

        Before using the ensemble gradient to train the neural network, which has $1021$ parameters, it is validated using a simplified one parameter turbulence model. 
        In this second validation test, the gradient of the cost function for the one parameter model is calculated for a range of values of the parameter and compared between the different methods.
        The model consists of $g^{(1)}=-C_\mu$ treated as a constant and the true value is taken as $C_\mu=0.09$. 
        The gradient is evaluated for a range of values of $C_\mu$ using both the ensemble and adjoint methods. 
        The ensemble sensitivities are approximated with $50$ samples.
        The results are shown in Figure~\ref{fig:comp_gradient}, and it can be seen that although they are noisy and have different magnitude the ensemble gradients result in the same sign and same zero as from the adjoint in the search region near the true value. 
        The ensemble gradients can be seen to be noisy but still were sufficient for correctly training the channel case (see Section~\ref{sec:results:channel}).
        This observation further confirms that the ensemble gradient can be used for training turbulence models. 

    \subsection{Learning a linear model}
    \label{sec:results:channel}
\begin{figure}[!htb]
            \centering
            \includegraphics[width=\textwidth]{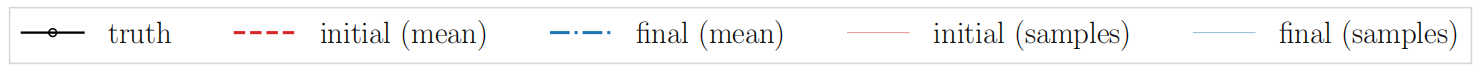} \subfloat[Velocity]{ \raisebox{-.5\height}{\includegraphics[width=0.45\textwidth]{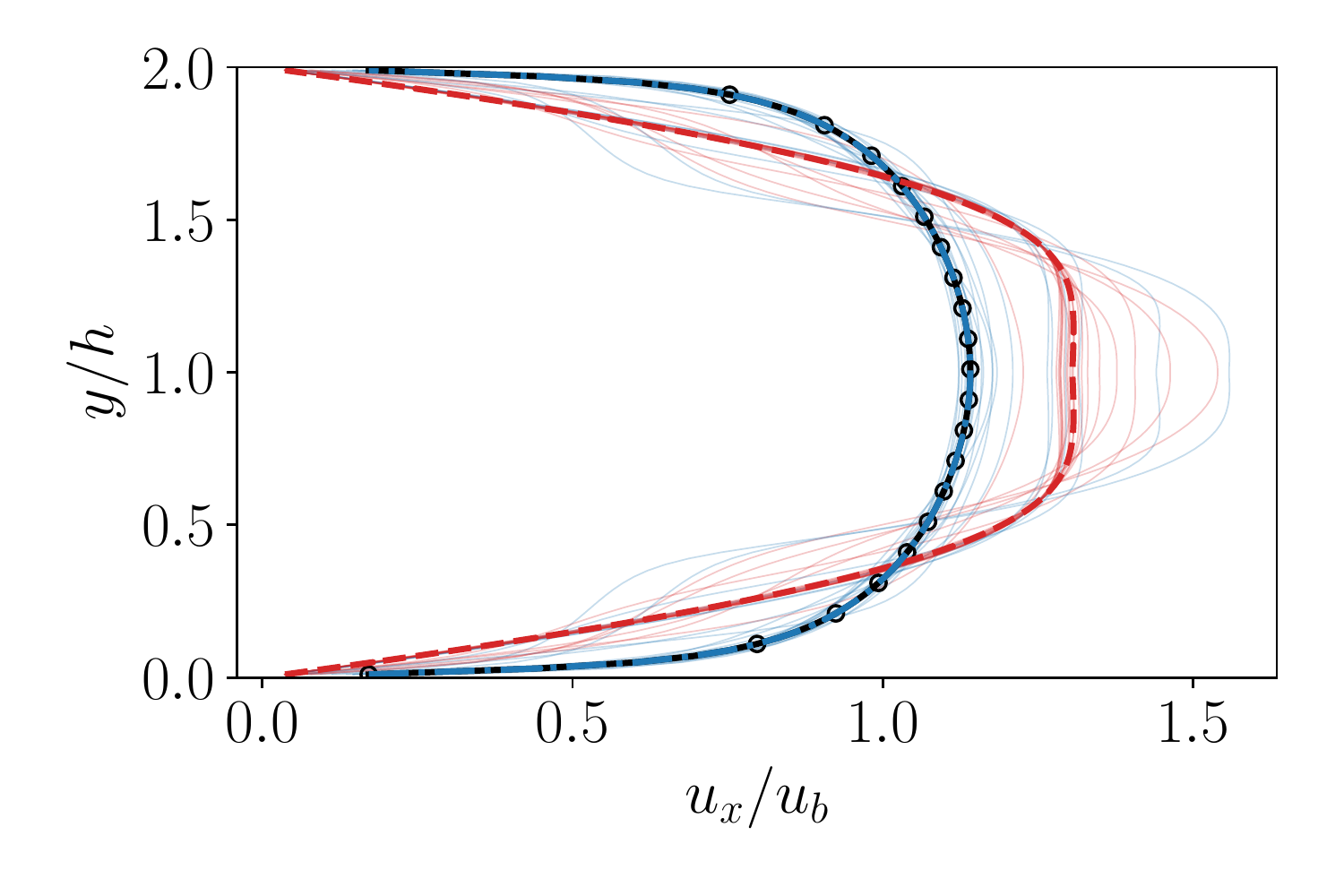}}} \hfill \subfloat[Function $g^{(1)}$]{ \raisebox{-.5\height}{\includegraphics[width=0.45\textwidth]{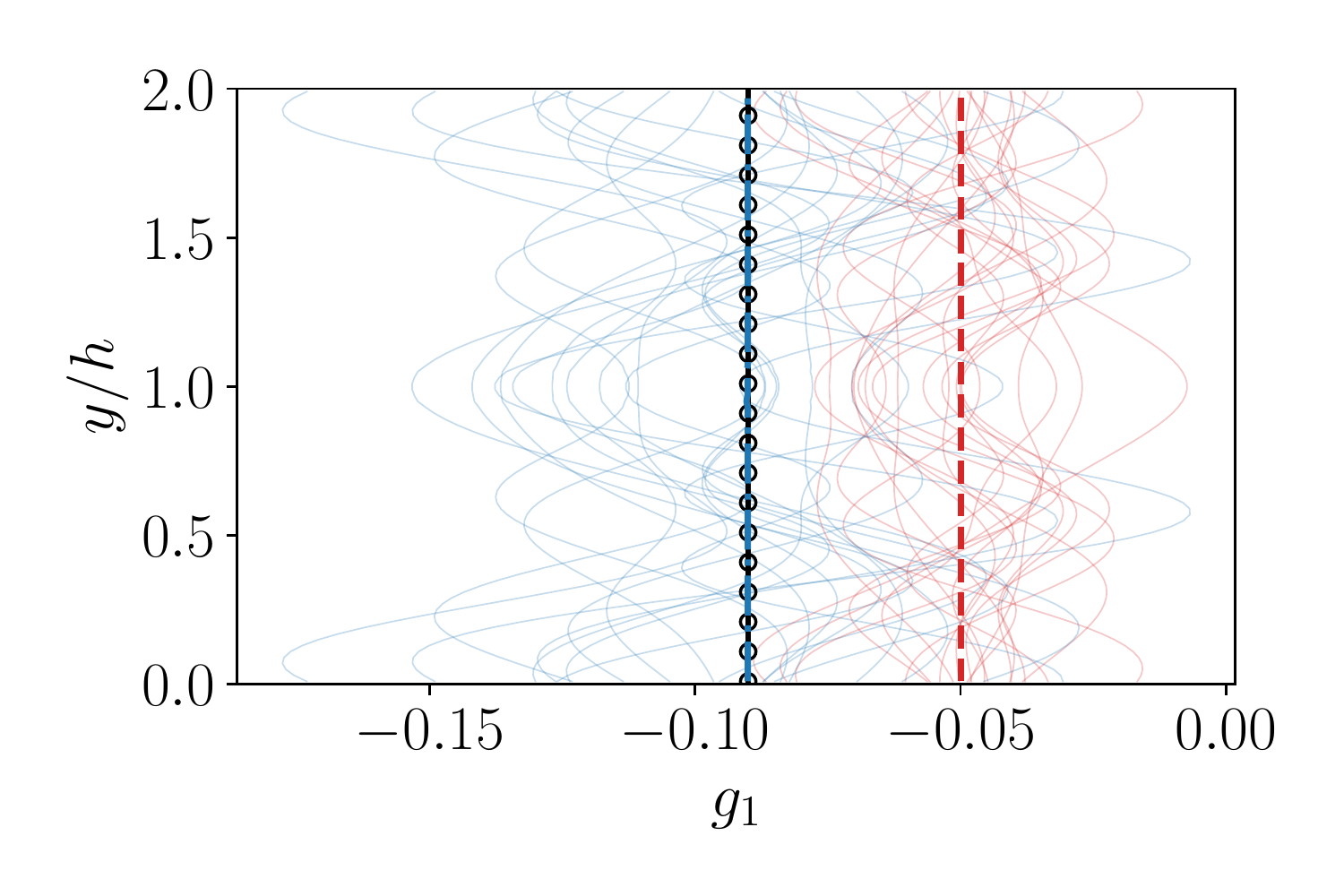}}}
            \caption{Results of learning a linear model from velocity data of the channel flow. The samples are used to estimate the gradient.
            The wall normal coordinate is indicated by $y$ with $y=h$ the center of the channel and $y = 0$ and $2h$ the bottom and upper walls, respectively. 
          }
            \label{fig:results_channel}
        \end{figure}

        The full-field velocity in the channel case is used to learn the linear turbulence model. 
        The neural network has one input $\theta_1$ and one output $g^{(1)}$ with $1,021$ parameters (trainable weights).
        The true solution is simply $g^{(1)}=-0.09$ which is constant and has no dependence on the input, but this is not enforced and must be learned.  
        The neural network is pre-trained to a constant $g^{(1)}=-0.05$ rather than to $g^{(1)}=0.0$ used in~\cite{michelen2020Data} since the later would result in samples with $g^{(1)}>0$ which are nonphysical and tend to result in diverging RANS simulations. 
        The network is then trained using the ADAM algorithm with default parameters. 
        For the gradient, the projection ensemble method presented in Section~\ref{sec:method:projection} is used with $20$ samples. 
        Figure~\ref{fig:results_channel} shows the results of the training including the initial and final samples used to estimate the gradient. 
        The trained model not only results in the correct velocity but learns the true underlying model for $g^{(1)}$.
        
    \subsection{Learning a quadratic model}
    \label{sec:results:duct}
\begin{figure}
            \centering
            \begin{tabular}{ccccl} 
                & Initial & Final & Truth & \\
                \rotatebox[origin=c]{90}{$u_x/u_b$} & 
                \raisebox{-.5\height}{\includegraphics[scale=0.75, trim=10 0 10 0, clip]{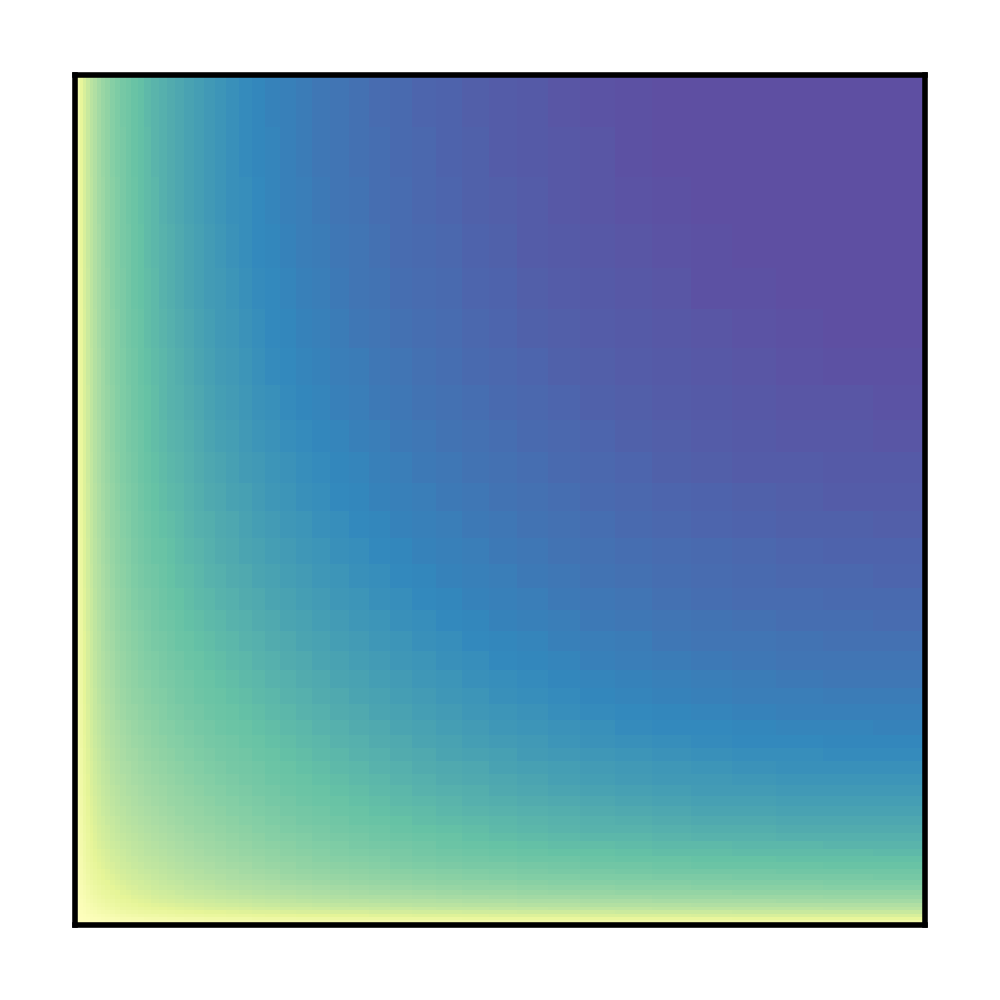}} &
                \raisebox{-.5\height}{\includegraphics[scale=0.75, trim=10 0 10 0, clip]{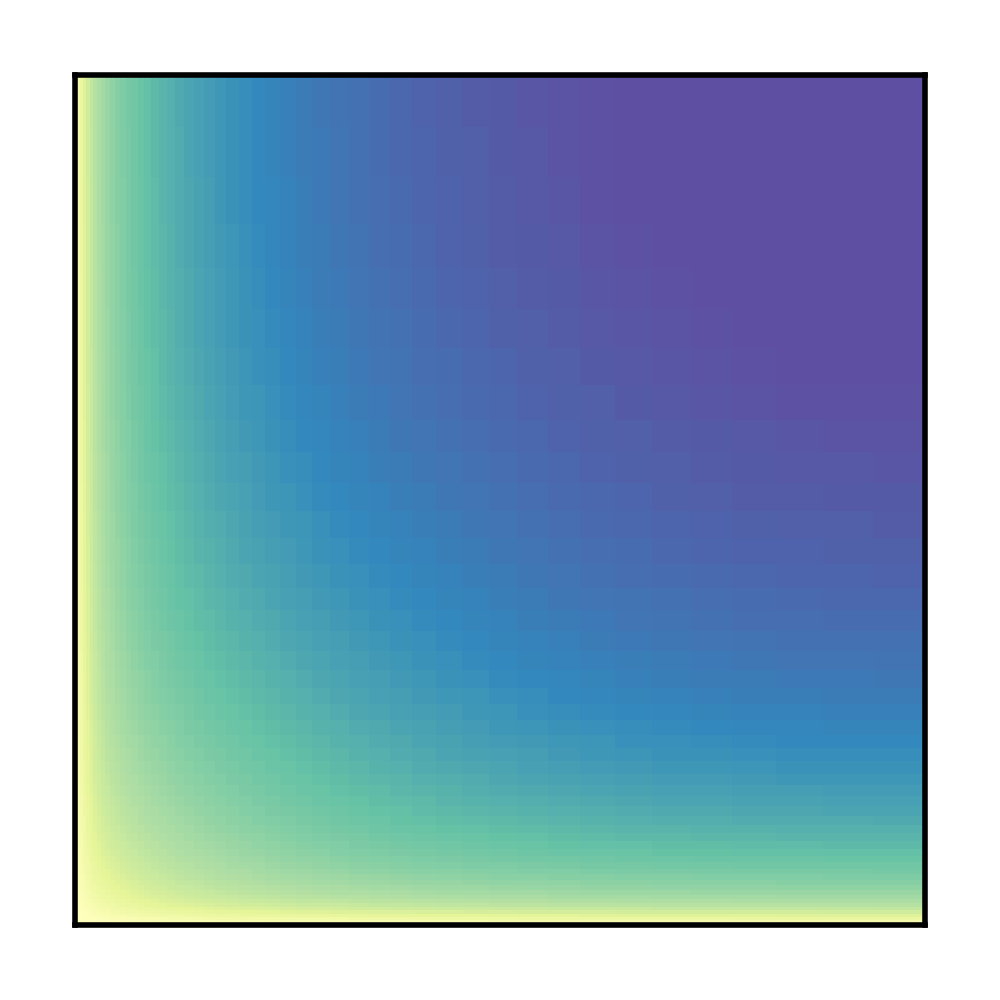}} &
                \raisebox{-.5\height}{\includegraphics[scale=0.75, trim=10 0 10 0, clip]{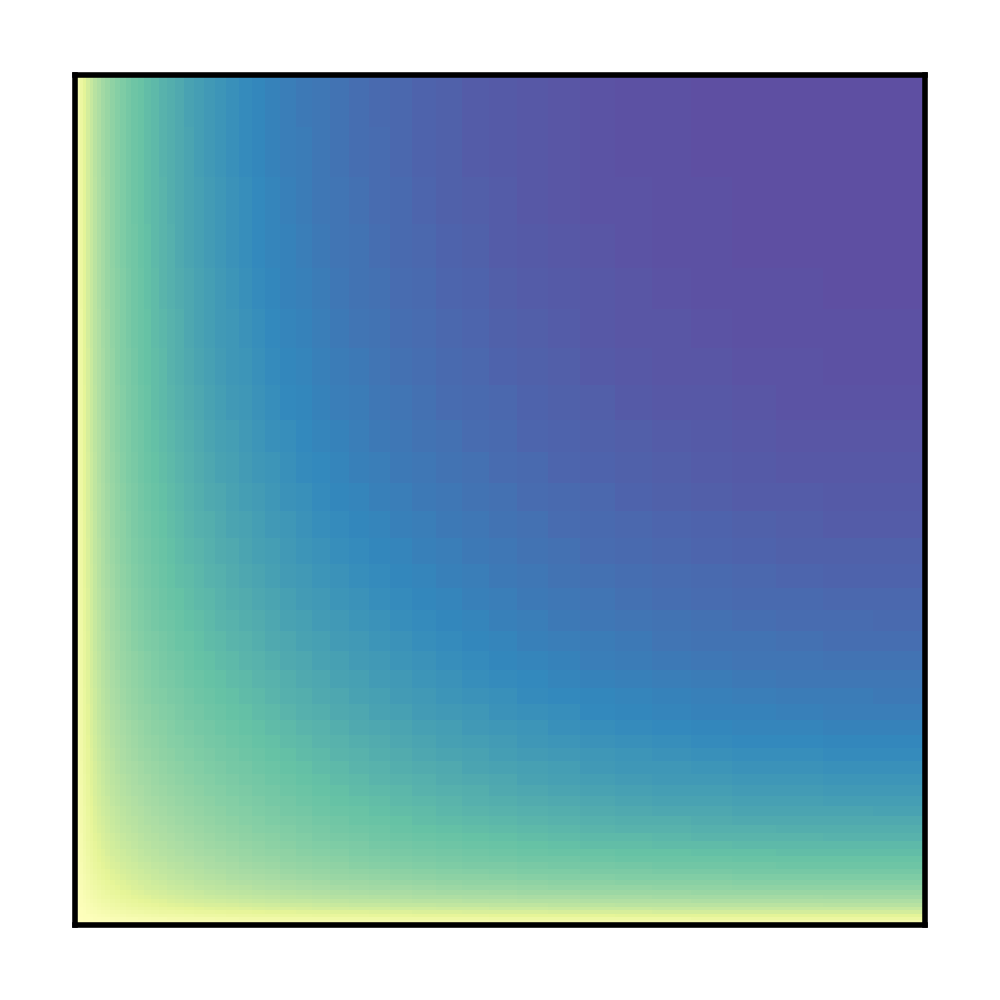}} & 
                \raisebox{-.5\height}{\includegraphics{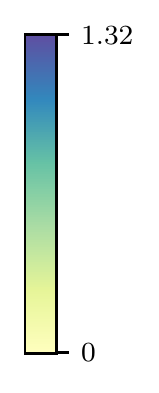}}
                \\
                \rotatebox[origin=c]{90}{$u_y/u_b\times10^{3}$} & 
                \raisebox{-.5\height}{\includegraphics[scale=0.75, trim=10 0 10 0, clip]{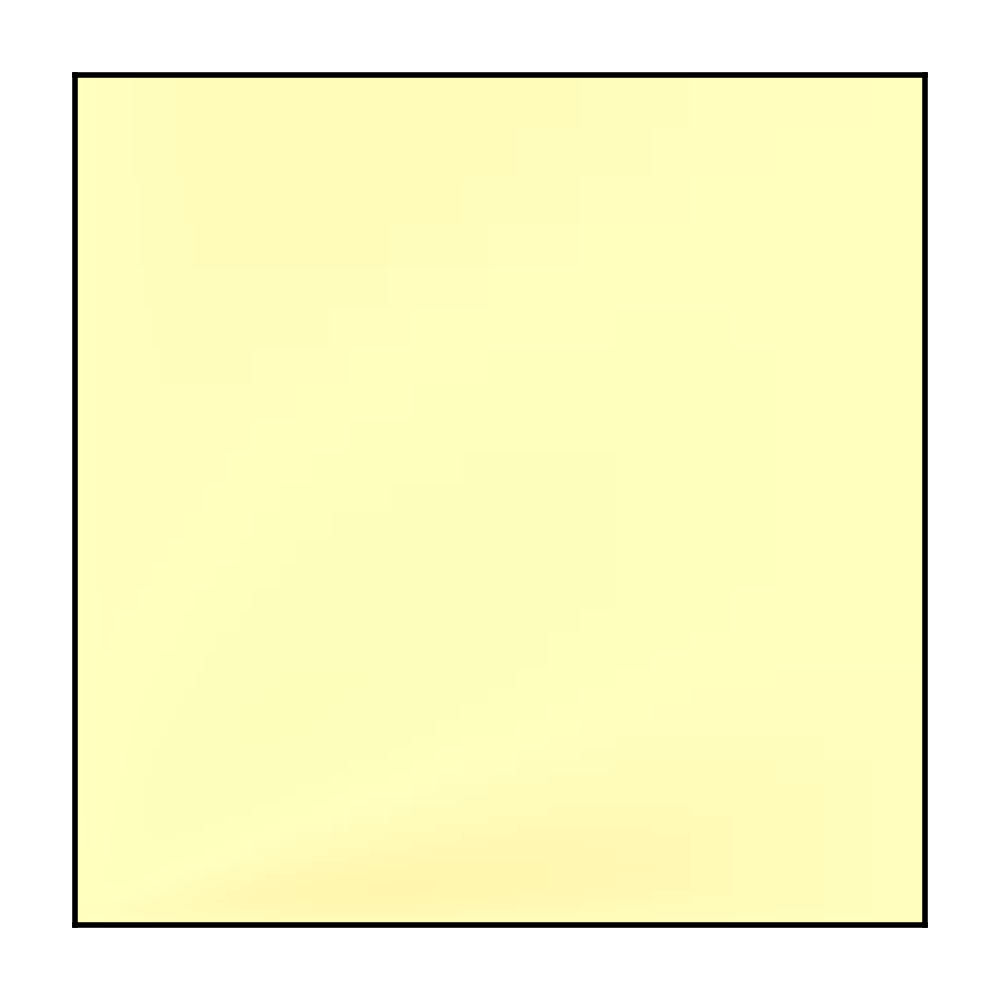}} &
                \raisebox{-.5\height}{\includegraphics[scale=0.75, trim=10 0 10 0, clip]{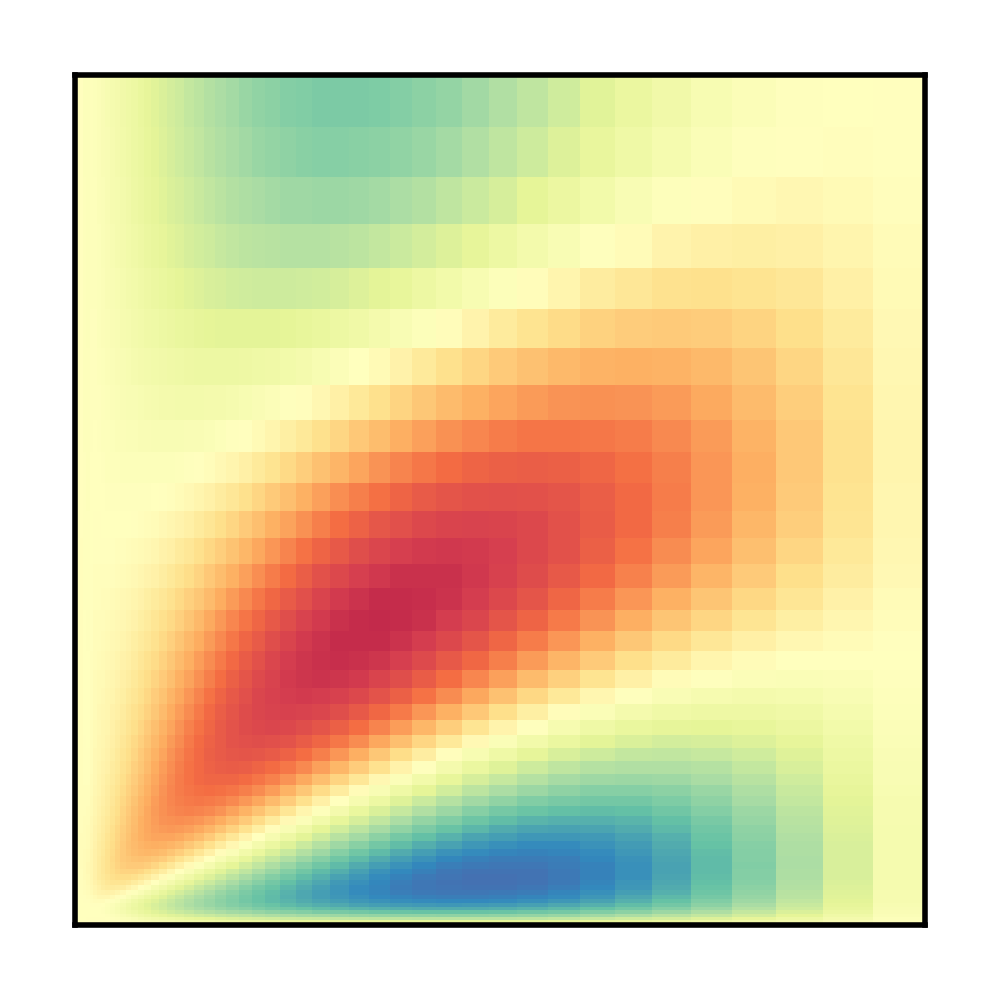}} &
                \raisebox{-.5\height}{\includegraphics[scale=0.75, trim=10 0 10 0, clip]{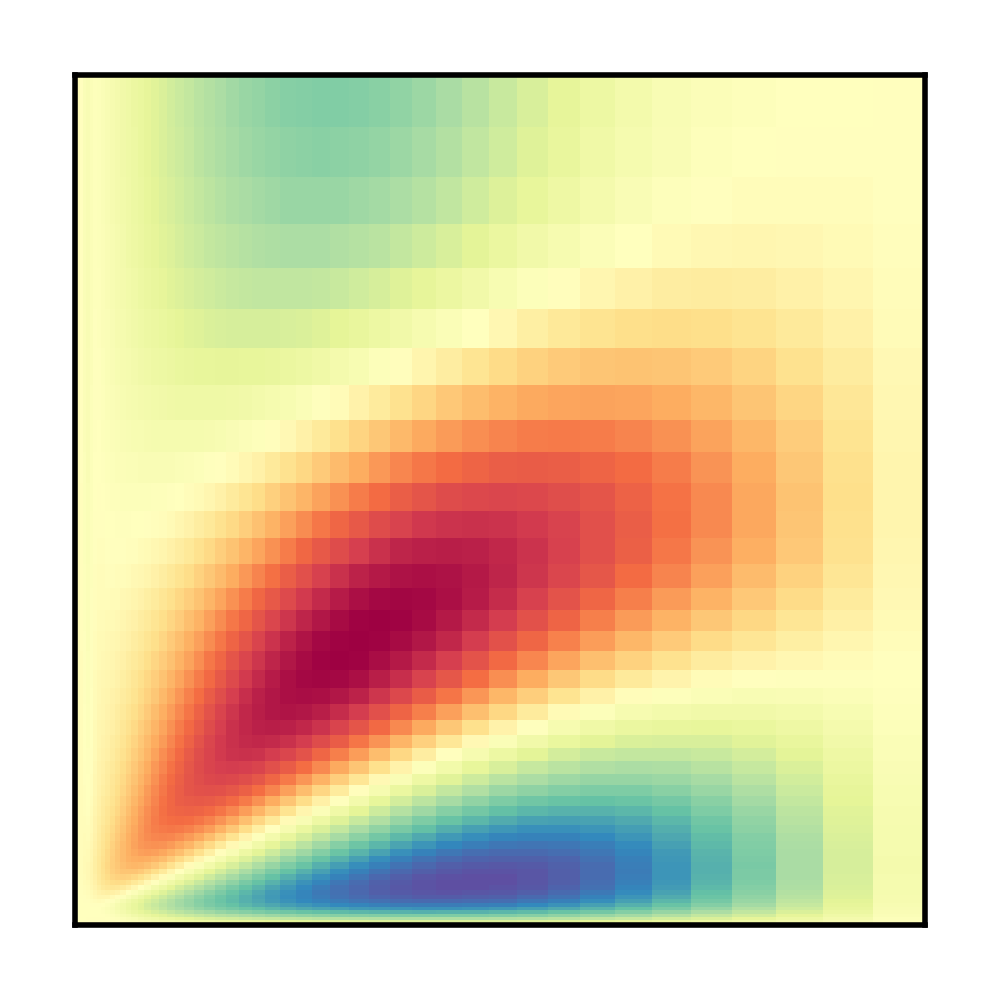}} & 
                \raisebox{-.5\height}{\includegraphics{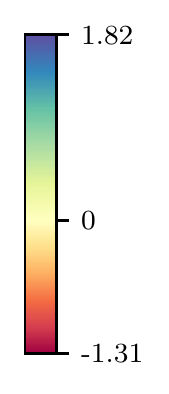}}
                \\
            \end{tabular}
            \caption{Velocity results of learning a NLEVM from full field velocity measurements in flow through a square duct. The axial ($u_x$) and in-plane ($u_y$) velocities are shown. The $u_z$ component is the reflection of $u_y$ along the diagonal through the center channel and is therefor not shown.}
            \label{fig:duct_U}
        \end{figure}

        The second test case consists of using the full velocity field in a flow in a square duct to learn a quadratic turbulence model. 
        The non-linear model is the Shih quadratic $k$--$\varepsilon$~\cite{shih1993realizable} given by 
        \begin{equation}
            \begin{array}{l}
                \displaystyle g^{(1)}(\theta_1, \theta_2) = \frac{-2/3}{1.25 + \sqrt{2\theta_1} + 0.9 \sqrt{-2\theta_2}}, \\[10pt]
                \displaystyle g^{(2)}(\theta_1, \theta_2) = \frac{7.5}{1000 + (\sqrt{2\theta_1})^3}, \\[10pt]
                \displaystyle g^{(3)}(\theta_1, \theta_2) = \frac{1.5}{1000 + (\sqrt{2\theta_1})^3}, \\[10pt]
                \displaystyle g^{(4)}(\theta_1, \theta_2) = \frac{-9.5}{1000 + (\sqrt{2\theta_1})^3}. 
            \end{array}
            \label{eq:shih_final}
        \end{equation}
        However, for the square duct case only $g^{(1)}$ and the combination $g^{(2)}-0.5g^{(3)}+0.5g^{(4)}$ affect the velocity, and $\theta_1\approx-\theta_2$. 
        Therefor a neural network with one input and two outputs is used. 
        The neural network is pre-trained to the linear model, $g^{(1)}=-0.09$ and $g^{(2)}=0$. 
        The training is first done with a small learning rate of $10^{-5}$ for initial stability of the optimization scheme, and later increased to the default value of $0.001$. 
        The gradient is obtained with the projection ensemble gradient using $20$ samples. 
        Figure~\ref{fig:duct_U} shows the velocity prediction of the trained model. 
        The trained model predicts the in-plane velocity, which the linear models fail to predict. 
        
        The trained model did not learn the underlying Shih quadratic model. 
        This is because only a small modification of the coefficients is needed to achieve improved velocity predictions. 
        Consequently, the sensitivity of the velocity to the underlying model becomes very small. 
        It was noted in~\cite{michelen2020Data} that learning the correct velocity with the adjoint gradient required only a few tens of training steps, while getting a better agreement in the underlying model took one to two orders of magnitude more steps.
        Unlike the adjoint, the ensemble gradient here represents only an approximation of the gradient, which is not accurate enough to continue training once the sensitivity of velocity to the underlying model becomes small. 
        
\section{Conclusion}
\label{sec:conclusion}
    In this paper an ensemble approximation of the gradient was used to obtain the sensitivities of the RANS equations during training of a data-driven turbulence model with indirect observations. 
    Through the use of a simple one parameter validation case it was shown that although the different ensemble approximations and the adjoint produce different gradients, they share the same zero-gradient location, sign, and qualitative behavior. 
    This case consisted of learning the scalar $C_\mu$ coefficient in a linear turbulence model from observations of synthetic velocity in a turbulent channel flow. 
    The same channel flow velocity was used to successfully train a deep neural network to learn a linear model.
    A second test case consisted of using synthetic velocity observations from flow in square duct using a non-linear model. 
    The learned model was able to predict the velocities, including the in-plane secondary velocity, but did not learn the correct underlying turbulence model. 
    This is due to the gradient approximation not being accurate enough to continue training once the sensitivity of velocity becomes small. 
    This is not expected to be an issue, however, when training more realistic models with a wide range of flows rather than with a single flow where the peculiarities of a single flow would have less weight.

    The present methodology can be used for any case where a neural network (or other differential machine learning models that provides its own gradients) is trained using quantities that require propagating the output of the network through another forward model. 
    The use of the ensemble gradient requires multiple evaluations of the forward model (e.g. RANS equations) at each step, but allows for it to be treated as a black box model. 
    On the other hand deriving and implementing the continuous or discrete adjoint can require significant overhead. 
    It is also noted that the ensemble gradient does not require the cost function to be locally differentiable as long there are clear global trends. 
    For these reasons, in some cases the approximate ensemble gradient might be preferable to the exact gradient from the adjoint method.

\section*{Acknowledgements}
    Michelén Ströfer is supported by the U.S. Air Force under agreement number FA865019-2-2204 while performing this research.
    The U.S. Government is authorized to reproduce and distribute reprints for Governmental purposes notwithstanding any copyright notation thereon.

\appendix

\section{Regularized projection}
\label{app:reg}
    The projection of the state $\xstate$ onto the space spanned by the sample discrepancies, described in Section~\ref{sec:method:projection}, is equivalent to minimization of the following objective function on $\beta$:
    \begin{equation}
        J_b =  \frac{1}{2} \lVert \Delta_{\xstate}\beta - (\xstate - \bar{\xstate}) \rVert^2_W. 
    \end{equation}
    We regularize the problem with Tikhonov regularization as
    \begin{equation}
        \argmin_\beta J_b(\beta) = \frac{1}{2} \lVert \Delta_{\xstate}\beta - (\xstate - \bar{\xstate}) \rVert^2_W + \lambda \frac{1}{2} \lVert \beta \rVert^2, 
    \end{equation}
    where $\lambda$ is the regularization strength parameter and can be chosen as small as possible while still providing the desired regularization.
    The derivative of the objective function with respect to $\beta$ can be formulated as
    \begin{equation}
        \nabla_\beta J_b =  \Delta_\xstate^\top W (\Delta_{\xstate}\beta - (\xstate - \bar{\xstate})) + \lambda \beta.
    \end{equation}
    By setting the derivative equals to zero and solving for $\beta$, we have
    \begin{equation}
        \beta = (\Delta_\xstate^\top W \Delta_{\xstate} + \lambda \mathsf{I})^{-1} \Delta_\xstate^\top W (\xstate - \bar{\xstate}).
    \end{equation}

    \begin{figure}[!htb]
        \centering
        \subfloat[without regularization]{        
        \includegraphics[width=0.45\linewidth]{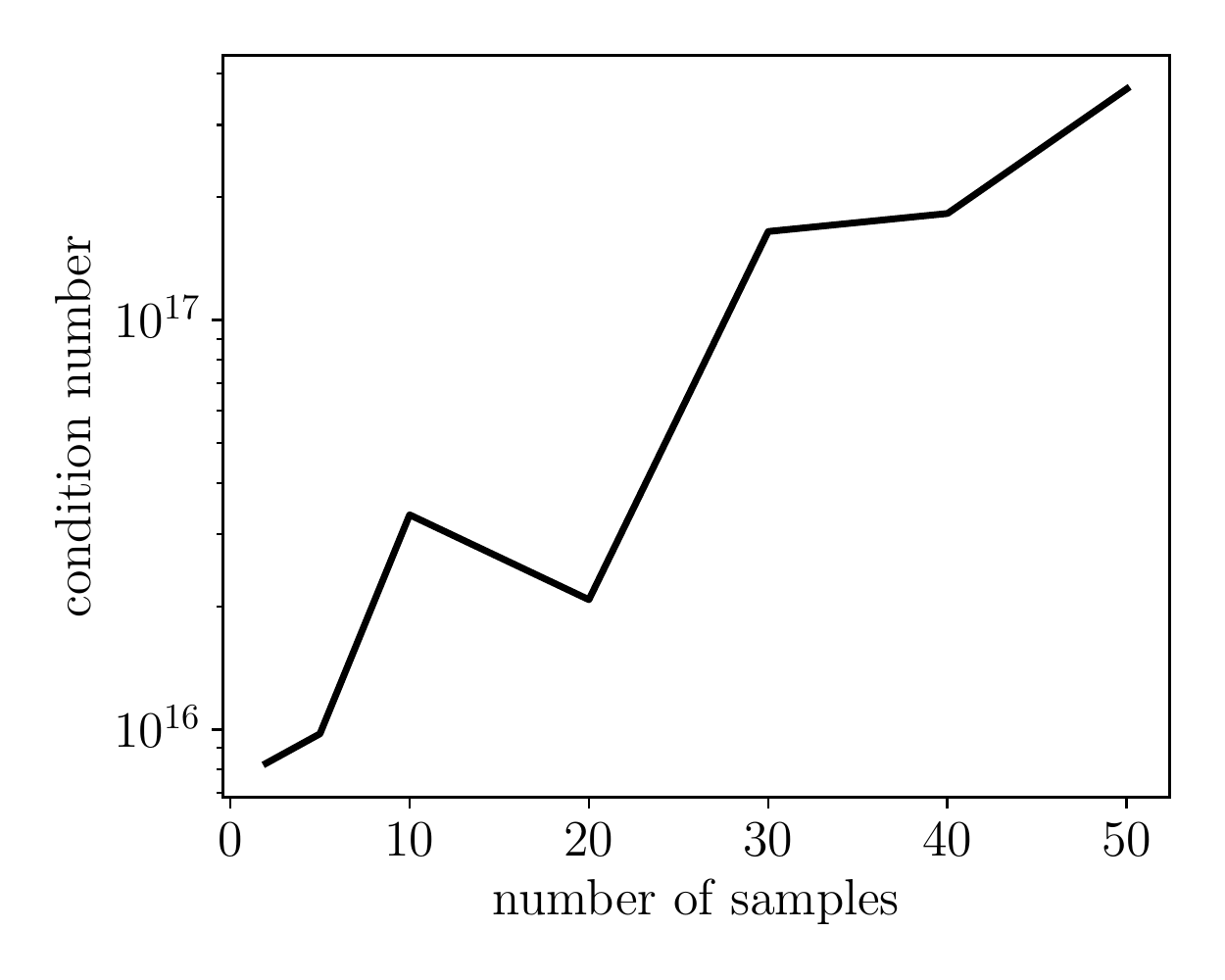}}
        \subfloat[with regularization]{        
        \includegraphics[width=0.45\linewidth]{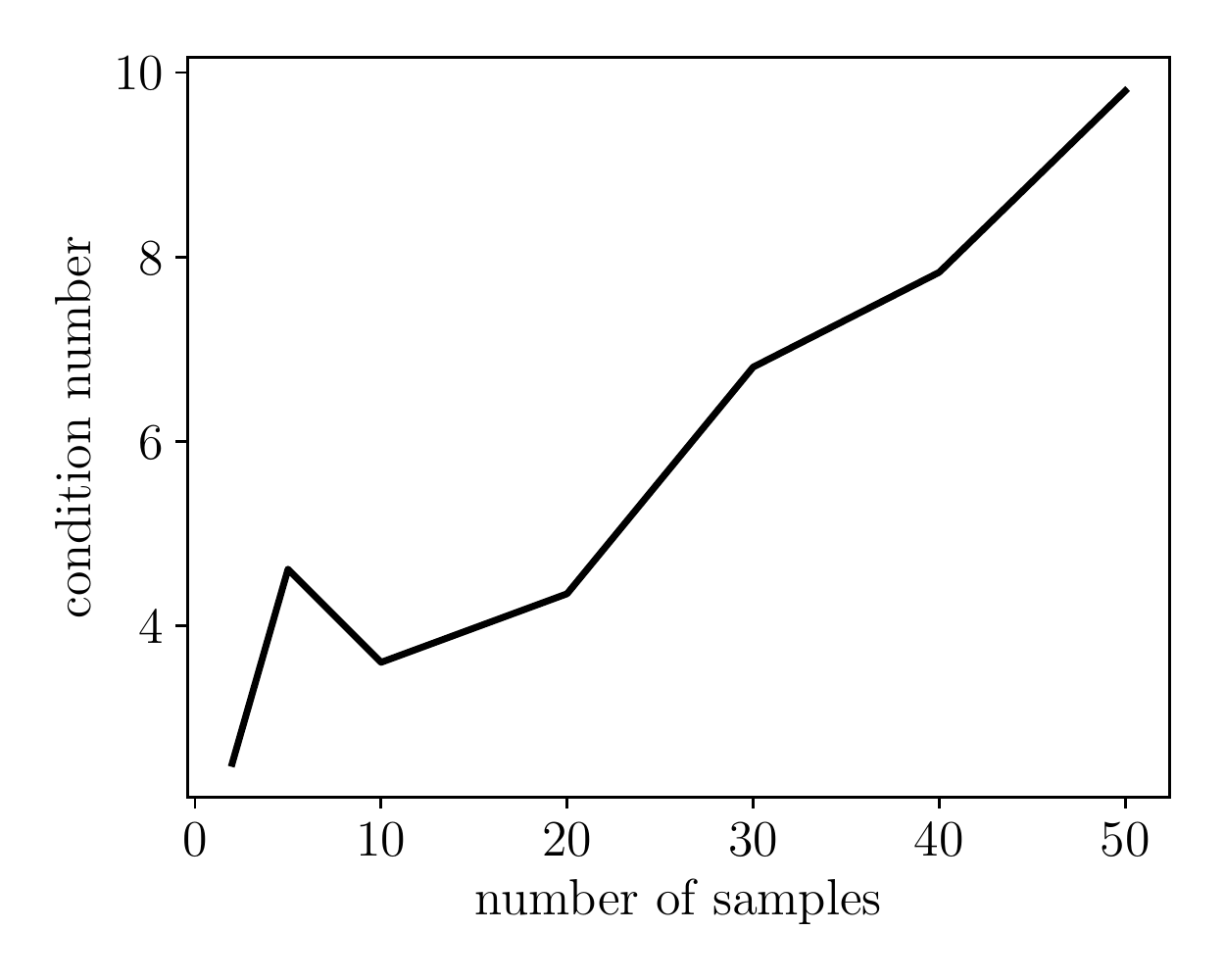}}
        \caption{Condition number of the matrix $\Delta_\xstate^\top W \Delta_{\xstate}$ for the channel case in Section~\ref{sec:results:channel} with and without regularization. The condition number is shown for ensembles with different number of samples.}
        \label{fig:condnum}
    \end{figure}
    
    The effectiveness of the regularization is demonstrated by looking at the condition number of the matrix to be inverted. 
    The condition number of $\Delta_\xstate^\top W \Delta_{\xstate}$ with and without regularization is presented in Fig.~\ref{fig:condnum} for the channel case in Section~\ref{sec:results:channel}. 
    The condition number if shown for ensembles with different number of samples and a regularization parameter of $\lambda=10^{-8}$ is used. 
    It can be seen that without regularization the matrix~$\Delta_\xstate^\top W \Delta_{\xstate}$ has very large condition number and is ill-conditioned even with only $2$ samples.
    With regularization, the matrix can be well conditioned with small condition numbers.

\section{State covariance as a preconditioner}
\label{app:premultiply}
    The preconditioned gradient descent in Equation~\eqref{eq:gradDescentP} represents a steepest descent on a discrete state space with inner product defined by the state covariance matrix. 
    That is, the direction of steepest descent depends on the choice of norm in the state space~\cite{tarantola2005inverse_ch6}. 
    The norm defines an infinitesimal circle and the steepest descent direction is towards the point in the circle that takes the minimum value of the objective function. 
    In general for L2 norms the steepest direction is only aligned with the gradient direction when the matrix defining the inner product is the identity matrix. 
    Otherwise, for an inner product $<a,b>_P = a^\top P b$ the steepest direction is simply $P$ times the gradient direction, as in Equation~\eqref{eq:gradDescentP}. 
    The use of the steepest direction can be seen as preconditioner or a regularization that results in smoother gradient approximation. 
    
    Alternatively, the use of the state covariance as a preconditioner can be derived from the quasi-Newton method, with some rough approximation, and a regularized cost function. 
    The regularized cost function is 
    \begin{equation}
        \tilde{J}(\xstate_n) = \frac{1}{2}\lVert \Hout(\xstate_n)-\yobs \rVert^2_{\Ccov_\yobs^{-1}} + \frac{1}{2}\lVert \xstate_n-\xstate_{n-1}\rVert^2_{\Ccov_\xstate^{-1}}. \label{eq:costReg}
    \end{equation}
    It has also been shown that, for the same regularized cost function, the continuous-time limit of the ensemble Kalman inversion behaves as the preconditioned gradient descent in Equation~\eqref{eq:gradDescentP}. 
    These two cases are summarized next. 
    
    \subsection{From quasi-Newton method}
        A typical way to derive a linear preconditioner for a gradient descent problem is using an approximation of the operator appearing in the quasi-newton algorithm. 
        Often, even a crude approximation works well~\cite{tarantola2005inverse_ch6}. 
        The state covariance as the preconditioner can be obtained in this manner with some approximations~\cite{nwaozo2006dynamic}.
        Using the regularized cost function in Equation~\eqref{eq:costReg}, the gradient and Hessian are 
        \begin{align}
            \nabla\tilde{J}(\xstate_n) &= (\Hout^\prime(\xstate_n))^\top \Ccov_\yobs^{-1}\left(\Hout(\xstate_n)-\yobs\right) + \Ccov_{\xstate_n}^{-1}(\xstate_n-\xstate_{n-1}), \\
            \nabla^2\tilde{J}(\xstate_n) &= \Ccov_\xstate^{-1} + (\Hout^\prime(\xstate_n))^\top \Ccov_\yobs^{-1}\Hout^\prime(\xstate_n) + (\Hout^{\prime\prime}(\xstate_n))^\top \Ccov_\yobs^{-1}\left(\Hout(\xstate_n)-\yobs\right),
        \end{align}
        where here $\nabla^2$ is the outer product ($\nabla\otimes\nabla$) indicating the Hessian. 
        For the quasi-Newton method the second derivative term $\Hout^{\prime\prime}$ is dropped, resulting in
        \begin{equation}
            \nabla^2\tilde{J}(\xstate_n) \approx \Ccov_\xstate^{-1} + (\Hout^\prime(\xstate_n))^\top \Ccov_\yobs^{-1}\Hout^\prime(\xstate_n),
        \end{equation}
        and the update scheme is 
        \begin{equation}
            \xstate_{n+1} = \xstate_n + \alpha_n \left(\nabla^2\tilde{J}(\xstate_n)\right)^{-1}\nabla\tilde{J}(\xstate_n).
        \end{equation}
        A crude approximation is obtained by treating the term $G(\xstate)=\lVert\Hout(\xstate)-\yobs\rVert^2_{\Ccov_\yobs^{-1}}$ without consideration to its form. 
        The Hessian then becomes
        \begin{align}
        \begin{split}
            \nabla^2\tilde{J}(\xstate_n) &= \Ccov_\xstate^{-1} + G^{\prime\prime}, \\
            &\approx \Ccov_\xstate^{-1}.
        \end{split}
        \end{align}
        The update scheme becomes 
        \begin{equation}
            \xstate_{n+1} = \xstate_n + \alpha_n(\xstate_n-\xstate_{n-1}) + \alpha_n\Ccov_{\xstate}(\Hout^\prime(\xstate_n))^\top \Ccov_\yobs^{-1}\left(\Hout(\xstate_n)-\yobs\right), 
        \end{equation}
        which becomes the original update scheme preconditioned by the state covariance (Equation~\eqref{eq:gradDescentP}), when ignoring the previous step term $(\xstate_n-\xstate_{n-1})$.

    \subsection{From ensemble Kalman inversion}
        The ensemble Kalman inversion (EKI)~\cite{iglesias2013ensemble} is a method for general problem inversion based on iterative application of the ensemble Kalman filter (EnKF). 
        It has recently been used for model learning~\cite{kovachki2019ensemble,schneider2020ensemble, schneider2020learning} and adapted to be able to enforce arbitrary constraints~\cite{wu2019adding,schneider2020imposing, zhang2020regularized}. 
        In the EKI the state is augmented to include the observations $\Hout(\xstate)$ and the problem is reformulated as an artificial dynamics problem where all non-linearities are moved to the dynamic model. 
        The problem can be formulated as a linear EnKF problem on the augmented state, but can also be re-expressed in terms of the non-linear operators and the original state~\cite{michelen2020DAFI}. 
        The EKI solves the regularized inverse problem in Equation~\eqref{eq:costReg} implicitly. 
        The gradient-free update for each sample is given as
        \begin{align}
        \begin{split}
            \xstate^{(i)}_{n+1} &= \xstate^{(i)}_{n} + \Ccov_{\xstate_n\zout_n}\left( \Ccov_{\zout_n} + \Ccov_\yobs \right)^{-1} \left(\Hout(\xstate_n) - \yobs\right), \\
            &= \xstate^{(i)}_{n} + \frac{1}{N}\Delta_{\xstate_n}\Delta_{\zout_n}^\top\left( \frac{1}{N}\Delta_{\zout_n}\Delta_{\zout_n}^\top + \Ccov_\yobs \right)^{-1} \left(\Hout(\xstate_n) - \yobs\right),
        \end{split}
        \end{align}
        where $j$ indicates the sample index and the superscript $n$ indicates the iteration step.
        In the continuous time limit ($\frac 1N \xrightarrow[]{} 0$) the evolution of a sample in the EKI is~\cite{schillings2017analysis} 
        \begin{equation}
            \frac{d \xstate^{(i)}}{dt} = -\Ccov_{\xstate \zout}\Ccov_\yobs^{-1}\left(\Hout(\xstate) - \yobs\right).
        \end{equation}
        That is, the continuous time limit of the EKI has an update direction consistent with preconditioning the original gradient with the state covariance (Equation~\eqref{eq:gradDescentP}).



\bibliography{references}

\end{document}

%% file: figure_1.tikz
\begin{tikzpicture}
\newcommand{\fighalfwidth}{7} 
\newcommand{\fighalfheight}{3.1} 
\newcommand{\blockwidth}{6.15} 
\newcommand{\blockheight}{3.75} 
\newcommand{\blockhspacing}{5} 
\newcommand{\txtys}{-3} 
\newcommand{\ysleftdraw}{-2} 
\newcommand{\ysrightdraw}{-2} 
\tikzset{
    boxstyle/.style={rounded corners=10, Black!75, thick,},
}
\newcommand{\nodesep}{6} 
\newcommand{\layersep}{10}
\newcommand{\nodesize}{4} 
\newcommand{\nhiddenlayers}{3} 
\newcommand{\nhiddennodes}{5}
\newcommand{\nionodes}{2}
\tikzset{
    nn_layout/.style={node distance=\nodesep mm and \layersep mm},
    nn_node/.style={circle, minimum size=\nodesize mm, draw=black!75, thin},
    nn_hnode/.style={nn_node, fill=black!5},  
    nn_inode/.style={nn_node, fill=SpringGreen!50},
    nn_onode/.style={nn_node, fill=Apricot!50},
    nn_connection/.style={very thin, draw=black!75},
}
\newcommand{\arrowlength}{0.47*\blockwidth} 
\newcommand{\arrowheight}{1.5} 
\newcommand{\arrowhead}{1.5} 
\newcommand{\arrowvspace}{2*\arrowhead} 
\tikzset{
    arrowfwdstyle/.style={draw=Black!75, thick, single arrow, inner sep = \arrowheight mm, single arrow head extend= \arrowhead mm, text width = \arrowlength cm, align=center, shading=axis, left color=SpringGreen!20, right color=White}, 
    arrowbwdstyle/.style={draw=Black!75, thick, single arrow, inner sep = \arrowheight mm, single arrow head extend= \arrowhead mm, text width = \arrowlength cm, align=center, shading=axis, right color=Apricot!20, left color=White, shape border rotate=180}, 
}
\newcommand{\varhspace}{2} 
\newcommand{\varvspace}{0} 
\newcommand{\Layer}[4]{
    \pgfmathtruncatemacro{\ys}{(#3-2)*(\nodesep/2)} 
    \pgfmathtruncatemacro{\xs}{(#2+0.5)*\layersep} 
    \begin{scope}[yshift=\ys mm]
        \coordinate (#1_1) at (\xs mm, 0);
        \node [#4] (#1_c_1) at (#1_1) {};
        \foreach \y in {2,...,#3}
        {
            \pgfmathtruncatemacro{\yp}{\y-1}
            \coordinate [below=of #1_\yp] (#1_\y);
            \node [#4] (#1_c_\y) at (#1_\y) {};
        }
    \end{scope}
}
\newcommand{\FullyConnect}[4]{
    \foreach \x in {1,...,#2}{
        \foreach \y in {1,...,#4}
        {
            \draw [nn_connection] (#1_c_\x.east) -- (#3_c_\y.west);
        }
    }
}
\newcommand{\neuralnet}{
    \begin{scope}[nn_layout, xshift=-(\nhiddenlayers+2)*0.5*\layersep mm, yshift=\nodesep*0.5 mm]
        \Layer{input}{0}{\nionodes}{nn_inode}
        \foreach \xx in {1,...,\nhiddenlayers}{
            \Layer{hidden_\xx}{\xx}{\nhiddennodes}{nn_hnode}
        }
        \Layer{output}{\nhiddenlayers+1}{\nionodes}{nn_onode}
        \FullyConnect{input}{\nionodes}{hidden_1}{\nhiddennodes}
        \pgfmathtruncatemacro{\yh}{\nhiddenlayers-1}
        \foreach \yy in {1,...,\yh}{
            \pgfmathtruncatemacro{\ypo}{\yy+1}
            \FullyConnect{hidden_\yy}{\nhiddennodes}{hidden_\ypo}{\nhiddennodes}
        }
        \FullyConnect{hidden_\nhiddenlayers}{\nhiddennodes}{output}{\nionodes}
    \end{scope}
}
\pgfmathtruncatemacro{\lblockxshift}{\blockhspacing+\blockwidth*10} 
\pgfmathtruncatemacro{\blockyshift}{\fighalfheight*10-\blockheight*10}
\pgfmathtruncatemacro{\xs}{\blockwidth*10/2}
\pgfmathtruncatemacro{\ystxt}{\blockheight*10+\txtys}
\pgfmathtruncatemacro{\ysrdraw}{\blockheight*10/2+\ysrightdraw}
\pgfmathtruncatemacro{\ysldraw}{\blockheight*10/2+\ysleftdraw}
\pgfmathtruncatemacro{\llmx}{-\blockhspacing-\blockwidth*10/2}
\pgfmathtruncatemacro{\rlmx}{\blockhspacing+\blockwidth*10/2}
\coordinate (leftlowermid) at (\llmx mm, \blockyshift mm);
\coordinate (rightlowermid) at (\rlmx mm, \blockyshift mm);
\begin{scope}[xshift=-\lblockxshift mm, yshift=\blockyshift mm]
    \draw[boxstyle] (0, 0) rectangle (\blockwidth cm, \blockheight cm) {};
    \begin{scope}[xshift=\xs mm, yshift=\ysldraw mm]
        \neuralnet
    \end{scope}
    \begin{scope}[xshift=\xs mm, yshift=\ystxt mm]
        \node [text width=\blockwidth cm, align=center] {Turbulence Model};
    \end{scope}
    \node [arrowfwdstyle, below=\arrowvspace mm of leftlowermid] (tlarrow) {Neural Network};
    \node [arrowbwdstyle, below=\arrowvspace mm of tlarrow] (blarrow) {Backpropagation};
\end{scope}
\pgfmathtruncatemacro{\ysarrow}{\blockyshift+\blockheight*10/2+\ysleftdraw}
\pgfmathtruncatemacro{\xspacearrow}{\blockhspacing-0.2}
\draw [-{latex}, Black, thick] (-\xspacearrow mm, \ysarrow mm) -- (\xspacearrow mm, \ysarrow mm);
\begin{scope}[xshift=\blockhspacing mm, yshift=\blockyshift mm]
    \draw[boxstyle] (0, 0) rectangle (\blockwidth cm, \blockheight cm) {};
    \begin{scope}[xshift=\xs mm, yshift=\ysrdraw mm]
        \node [text width=\blockwidth cm, align=center] (rans)
        { 
        $\begin{array}{c} 
        \boldsymbol{u\cdot\nabla u} - {\nu}\boldsymbol{\nabla^2u} + \boldsymbol{\nabla\cdot} \textcolor{black}{\reynoldstress} + \nabla p - \boldsymbol{s} = 0 \\[4pt]
        \boldsymbol{\nabla\cdot u} = 0 
        \end{array}$
        };
    \end{scope}
    \begin{scope}[xshift=\xs mm, yshift=\ystxt mm]
        \node [text width=\blockwidth cm, align=center] {Objective Function};
    \end{scope}
    \node [arrowfwdstyle, below=\arrowvspace mm of rightlowermid] (trarrow) {PDE};
    \node [arrowbwdstyle, below=\arrowvspace mm of trarrow, right color=red!60, left color=red!60] (brarrow) {PDE Sensitivity};
\end{scope}
\node [left = \varhspace mm of blarrow, align=center] (leftvarnode) {\Large$\frac{\partial J}{\partial \boldsymbol{w}}$};
\node [right = \varhspace mm of trarrow] (rightvarnode) {$J$};
\coordinate (leftvar) at (leftvarnode);
\coordinate (rightvar) at (rightvarnode);
\node [align=center] at ($(leftvar |- tlarrow.west)$) {$\boldsymbol{w}$};
\node at ($(tlarrow.east -| 0,0)$) {$\reynoldstress$};
\node [right = \varhspace mm of trarrow] {$J$};
\node [below = \varvspace mm of brarrow] {$\partial J/\partial \reynoldstress$};
\node [below = \varvspace mm of blarrow] {\small$\partial \reynoldstress/\partial \boldsymbol{w}$};
\end{tikzpicture}